*Astronomy & Astrophysics*

# In-orbit background of X-ray microcalorimeters and its effects on observations

S. Lotti[1], D. Cea[1], C. Macculi[1], T. Mineo[2], L. Natalucci[1], E. Perinati[3], L. Piro[1], M. Federici[1], and B. Martino[1]

[1] INAF-IAPS Roma, via fosso del cavaliere 100, 00133 Rome, Italy
e-mail: simone.lotti@iaps.inaf.it
[2] INAF-IASF Palermo, via Ugo la Malfa 153, 90146 Palermo, Italy
[3] IAAT – Institute für Astronomie und Astrophysik, Universität Tübingen, 72076 Tübingen, Germany

**ABSTRACT**

*Context.* Observations in the X-ray energy band are often limited by the background because of the low fluxes of typical sources. The background can easily be higher than the signal itself, and thus any mission with the scientific goal of observing faint and/or extended sources in the X-ray band must deal with the background problem. ESA has recently selected "the hot and energetic universe" as science theme for the second large-class mission in the Cosmic Vision science program, to be pursued with an advanced X-ray observatory to be launched in 2028, and at present *ATHENA* is the proposal that is most likely to be selected for this slot. The mission is aimed to place an X-ray telescope in the L2 orbit equipped with an X-ray Integral Field Unit (X-IFU) based on high spectral resolution transition-edge microcalorimeters, and has among its goals the detection and characterization of high-redshift active galactic nuclei (AGNs), cluster of galaxies and their outskirts, which is why great care must be taken to reduce the background impact on the detection/characterization of these sources.
*Aims.* The background is composed of a diffuse component and an internal particle component for any satellite operating in the X-ray band. We take as reference the X-ray IFU instrument that will be placed onboard the *ATHENA* mission to analyze both these components and their variability for different orbits, observational conditions and/or design choices. We also show how different background levels affect the instrumental performance, and the scientific results obtainable with the instrument in the best configuration. The X-IFU is a cryogenic X-ray spectrometer, based on a large array of 3840 transition-edge sensors (TES) of 250 µm side, with a spectral energy resolution of 2.5 eV in the 0.2 − 10 keV energy band, over a field of view of 5 × 5 arcmin$^2$, high count rate capability and a 5 arcsec angular resolution.
*Methods.* There are no experimental data about the background experienced by microcalorimeters in the L2 orbit, and thus the particle background levels were calculated by means of Monte Carlo simulations: we considered the original design configuration and an improved configuration aimed to reduce the unrejected background, and tested them in the L2 orbit and in the low Earth orbit, comparing the results with experimental data reported by other X-ray instruments. For the diffuse component, we used the background levels measured from a 1 sr region representative of typical high galactic latitude pointings and analyzed the variations expected with the different orbital conditions. To show the results obtainable with the improved configuration we simulated the observation of a faint, high-redshift, point source ($F_{[0.5-10 \text{ keV}]} \sim 6.4 \times 10^{-16}$ erg cm$^{-2}$ s$^{-1}$, $z = 3.7$), and of a hot galaxy cluster at $R_{200}$ ($S_{b[0.5-2\text{keV}]} = 8.61 \times 10^{-16}$ erg cm$^{-2}$ s$^{-1}$ arcmin$^{-2}$, $T = 6.6$ keV).
*Results.* First we confirm that implementing an active cryogenic anticoincidence reduces the particle background by an order of magnitude and brings it close to the required level. The implementation and test of several design solutions can reduce the particle background level by a further factor of 6 with respect to the original configuration. The residual background is dominated by secondary particles, and this component can be decreased by design solutions such as passive shielding with appropriate materials. The best background level achievable in the L2 orbit with the implementation of ad-hoc passive shielding for secondary particles is similar to that measured in the more favorable LEO environment without the passive shielding, allowing us to exploit the advantages of the L2 orbit. We define a reference model for the diffuse background and collect all the available information on its variation with epoch and pointing direction. With this background level the *ATHENA* mission with the X-IFU instrument is able to detect ∼4100 new obscured AGNs with $F > 6.4 \times 10^{-16}$ erg cm$^{-2}$ s$^{-1}$ during three years, to characterize cluster of galaxies with $S_b(0.5 − 2\text{keV}) > 9.4 \times 10^{-16}$ erg cm$^{-2}$ s$^{-1}$ sr$^{-1}$ on timescales of 50 ks (500 ks) with errors $< \sim 40\%$ ($< \sim 12\%$) on metallicity, $< \sim 16\%$ (∼4.8%) on temperature, ∼2.6% (∼0.72%) on the gas density, and several single-element abundances.
*Conclusions.* With respect to the original design the background level obtained significantly enhances the performance of the X-IFU. It allows us to improve the times needed for the detection and the characterization of high-redshift and/or faint Compton-thick AGNs and to reduce the limit flux of X-IFU, and thus to have a significant measurement of the fraction of these objects where current surveys are missing data. Moreover, the reduced background allows an improved characterization of diffuse and/or faint sources in terms of relative errors on the physical properties of the sources and observational times needed to obtain them.

## 1. Introduction

On 28 November 2013, "the hot and energetic Universe" science theme was selected for the second L-class mission in the ESA's Cosmic Vision science program, to be pursued with an advanced X-ray observatory. This mission, scheduled for 2028, will address two key questions: how does ordinary matter assemble into the large-scale structures that we see today, and how do black holes grow and shape the Universe? The favored candidate to answer these question is the *ATHENA* observatory class mission concept, which will be presented for the ESA call in 2014. *ATHENA* will be placed in orbit at L2, around the second Lagrangian point of the Sun-Earth system in a large halo orbit, with a semi-major axis amplitude of about 700 000 km and a period of approximately 180 days. The mission includes two focal plane detectors: a Wide Field Imager (WFI; Nandra et al. 2013), and the X-ray Integral Field Unit (X-IFU).

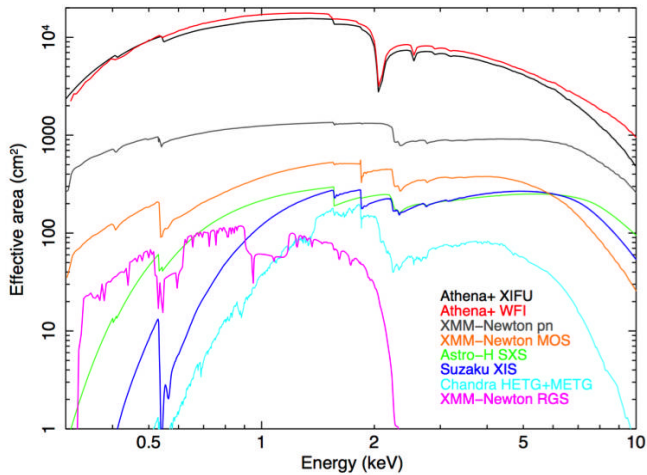

**Fig. 1.** Effective area as function of energy for *ATHENA* and other operational and planned missions.

In this paper we illustrate the background reduction achievable for a microcalorimeter-based instrument, using as a baseline the X-IFU technical specifications as presented in the strawman mission profile presented to ESA to meet the guidelines for "the hot and energetic Universe".

X-IFU is an array of 3840 transition-edge sensors (TES) 250 $\mu$m side, composed of Ti/Au sensors (the goal is to have Mo/Au TES because of their higher temperature stability) and 1 $\mu$m Cu and 4 $\mu$m Bi absorbers (goal: Au/Bi) that operates at cryogenic temperatures to achieve the high spectral resolution of $\sim$2 eV at 6 keV (Barret et al. 2013). The other focal plane instrument is the WFI. Its purpose is to provide X-ray images in the 0.1−15 keV energy range, with a large field of view (40 × 40 arcmin$^2$), excellent spatial and energy resolution (150 eV at 6 keV) and maximum count rate capabilities (>1 Crab) (Rau et al. 2013).

The *ATHENA* optics provide a unique combination of large area (2 m$^2$ at 1 keV, see Fig. 1), high angular resolution (5″ half energy width), and large field of view (40′ for the WFI and 5′ for X-IFU), thanks to the silicon pore optics (SPO) technology developed over the last decade (Willingale et al. 2013).

For any satellite operating in the X-ray band the background is composed of an internal particle component and a diffuse component. The former is generated by particles traveling through the spacecraft, releasing energy inside the detector, also creating swarms of secondary particles (mostly electrons) along the way. These secondaries can also reach the detector, and their flux is often anything but negligible in the soft X-ray band since it is hard to discriminate them using anticoincidence systems (Lotti et al. 2012). This can be a problem especially for "naked" devices such as microcalorimeters or back-illuminated CCDs, that do not present an inactive layer on the surface capable of absorbing such low-energy particles (Lotti et al. 2013). The latter component is the soft X-ray background (SXRB): a diffuse X-ray emission observed in every direction. At high energies it is generated by the unresolved emission from distant active galactic nuclei (AGNs), while at low energies ($\lesssim$1 keV) it is composed of several contributions, each with its own spatial and temporal variations, which makes it difficult to predict the exact level of background to be expected for any specific observation. Such a component is modeled using *XSPEC – version 12.8.0* as reported by (Mc Cammon et al. 2002) and described in Sect. 3.

In Sect. 2 we analyze the internal background induced by charged particles reaching the detector through the spacecraft in different orbital conditions and with different design choices: starting from the results obtained in Lotti et al. (2012; representative of the original design of the instrument, with background requirements derived from the previous Athena and IXO concepts), we show that an internal background level $\sim$6 times lower can be obtained for these detectors with clever design and material choices. We then explore the background dependence on different orbital conditions and compare the levels obtained with those recorded by existing instruments in different environments. In Sect. 3 we deal with the diffuse background. Differently from the internal component, the diffuse background is not affected by design choices in the detector neighborhood, so we define a model that will be chosen as reference for the X-IFU instrument and assess its expected variations for different pointing or epoch conditions. Finally, in Sect. 4, we show the effects of our improvements by simulating the observation of two representative frames indicated in the ESA call for mission concepts 2014 using the software *XSPEC*.

## 2. Internal particle background

There are no experimental data about the non-X-ray background level experienced by X-ray microcalorimeters in L2, and since *ATHENA* can be the first X-ray large emission placed there, we estimate it with detailed Monte Carlo simulations using the *Geant4.9.4* software[1]. The first background estimates have been extensively discussed by Lotti et al. (2012). Summarizing, as shown in Fig. 2a, given the satellite mass model without an anticoincidence detector (ACD), the detector would experience a particle background level of 3.1 cts cm$^{-2}$ s$^{-1}$, mainly induced by primary protons (80% of the total background). The insertion of the ACD and secondarily the use of pattern-recognition algorithms decreases the background by an order of magnitude to a level of 0.31 cts cm$^{-2}$ s$^{-1}$ in the 0.2−10 keV energy band, cutting the primary proton component and leaving a residual background rate induced mostly by secondary electrons. We refer to this background count rate as the "original" or "nominal" background value, since it was obtained in the official configuration foreseen for *ATHENA* before this work (Lotti et al. 2012). This performance is possible thanks to a solution developed by our team, which is adopting an active cryogenic ACD (Macculi et al. 2012, 2013, and references therein) to be placed underneath ($\sim$mm) the X-IFU main detector. This is a cryogenic microcalorimeter composed of a 2 × 2 pixel array made of large area silicon absorbers sensed by Ir TES. The main requirements of this ACD that we identified are a size of at least 18 × 18 mm$^2$ (divided among four identical pixels each 80 mm$^2$) to have sufficient geometrical rejection efficiency, a pulse rise time (which is used for the veto signal) lower or equal to that of the main detector, 30 $\mu$s, and a low-energy threshold for the detection of background particles of 20 keV. Prototypes with performance close or within the requirements have already been produced (Macculi et al. 2012, 2013).

Since this background level was still higher than the value of the scientific requirement of the mission (0.2 cts cm$^{-2}$ s$^{-1}$),

---
[1] http://geant4.web.cern.ch/geant4/

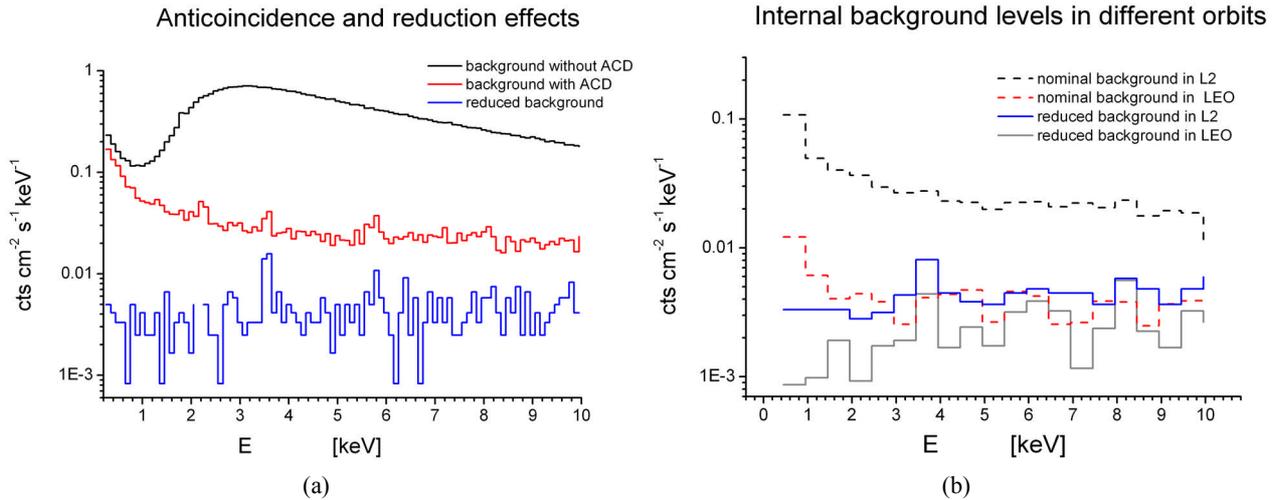

**Fig. 2. a)** Spectra (100 eV bins) of the background expected on X-IFU in all the cases analyzed. The black line is the expected background without an anticoincidence system, the red line is the level expected with an ACD, and the blue line is the final background level. **b)** Spectra (500 eV bins) expected on X-IFU in L2 and in LEO orbit, in the nominal configuration (dashed lines), and in the reduced configuration (continuous lines).

we searched for solutions to reduce the secondary-electron component that constituted 85% of the unrejected background and was mainly created in the internal cryostat wall. These solutions consist of

- Distance optimization: reducing the distance between the main detector and the ACD from 2 mm to 1 mm enhances the fraction of charged particles that cross the main detector that intercept the ACD as well, improving the geometrical rejection efficiency. The particles most affected are the primaries, because they possess enough energy to cross both detectors, while secondary particles are more likely to be absorbed entirely inside the main detector and are less effected. This reduced the nominal background by ~20%.
- Kapton coating: inserting a ~250 μm Kapton layer between X-IFU and the last surface seen by the detector. The coating absorbs the electrons produced in the cryostat innermost surface (Nb), generating a much lower electron flux toward the detector. This reduces the secondary electrons flux on the detector, and reduces the nominal background by ~70%.
- Kapton filter: moving one of the radiation filters (~100 nm thick) close to the detector surface to stop the secondary-electron flux toward the main detector, as we do with the Kapton coating. This way we can also exploit the trace left in the ACD by the primary particle to discriminate the secondary electrons produced inside the filter, which enhances the rejection efficiency for the secondary particles. However because of the very low thickness of the filter (100 nm), it is not as efficient as the coating and decreases the nominal background by ~17% (mostly below 1 keV). But it is not possible to increase the thickness of the filter since it is placed on the optical axis of the instrument and it would increase the X-ray absorption.

Combining all these solutions, we are able to cut the background by another factor of 6 (~83%), bringing us to the unrejected background level of 0.05 cts cm$^{-2}$ s$^{-1}$ reported here. We refer to this value as the "reduced background". This background is the official background level adopted for *ATHENA*.[2] All the background spectra considered so far are reported in Fig. 2a.

---

[2] Available from http://www.the-athena-X-ray-observatory.eu/

If we move from the L2 orbit to the LEO environment, the optimized design that allowed us to gain a factor of 6 in L2 this time in LEO reduces the background only by a factor of <2 (in fact as we can see in Fig. 2b we obtain similar results in LEO for any configuration). This is because the main contribution to the L2 background is given by the secondary electrons created by minimum ionizing particles (MIP) protons; in L2 we can cut this contribution with our design improvements, but in LEO most MIP (those with $E \lesssim 88$ keV, which represent ~90% of the CR protons in the worst case at the solar maximum) cannot reach the detector since they are blocked by the Earth magnetic field. Furthermore, while in L2 the unrejected background is almost completely induced by CR protons (Lotti et al. 2012), in LEO there are several other background components for which the design optimization is less effective. Our results for LEO agree with the background of ~0.05 cts cm$^{-2}$ s$^{-1}$ measured by the *Suzaku* microcalorimeter array (XRS; Perinati et al. 2012), which is the same kind of instrument placed in the same orbit.

As experienced by *XMM-Newton* and *Chandra*, low-energy solar protons can reach the focal plane focused by the optics. Their flux is driven by the optics effective area and in this case, to compute the proton rate at the focal plane, we use a ray-tracing simulator that includes the optics geometry, a model for the reflection of the protons from the mirror surface and of the scattering angles expected after the reflection. For *ATHENA* the ray-tracing simulations are currently being upgraded, and the only result available is the value obtained for IXO that reports a mean focused proton background of 0.01 cts cm$^{-2}$ s$^{-1}$ (Perinati et al. 2010).

### 2.1. Solar particle events and instrument dead time

In this section we estimate the total fraction of mission time where X-IFU will not be operative because of solar energetic particles. During solar events the flux of soft protons, usually negligible, can increase by several orders of magnitude. These particles can generate a count rate so high that the detectors do not have a stable working point. If the time spent out of the working point, that is the dead time (DT) of the instrument, is too high, the detector is shut down, and for X-IFU the DT threshold

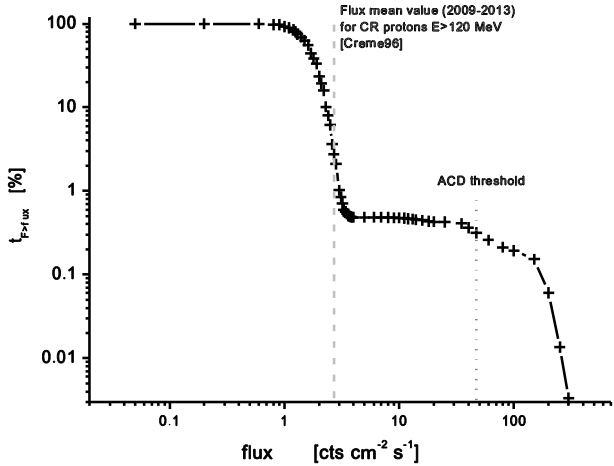

**Fig. 3.** Fraction of the mission time where the proton flux measured by SREM ($F$) is above a certain flux value. The proton fluxes ($E > 120$ MeV) are measured by the Solar Radiation Environment Monitor onboard Planck in the 2009−2013 time span. The flux drops rapidly around the mean value foreseen by Creme96 in the same timespan (vertical dashed line), while the second plateau (above $\sim 3$ cts cm$^{-2}$ s$^{-1}$) depicts the contribution of solar flares. The vertical dotted line shows the ACD threshold flux value where DT $> 5\%$.

**Table 1.** Background levels in CCDs, silicon drift detectors and microcalorimeters in different orbits.

| Mission | Detector name | Detector type | Orbit | Background [cts cm$^{-2}$ s$^{-1}$] |
|---|---|---|---|---|
| Suzaku | XIS | CCD (BI) | LEO | 0.048 |
| Suzaku | XIS | CCD (FI) | LEO | 0.016 |
| Suzaku | XRS | Microcalorimeters | LEO | 0.05 |
| Origin | X-IFU | Microcalorimeters | LEO | **0.043** |
| Simbol-X | LED* | Silicon drift | LEO | **0.002** |
| XMM | EPIC | pn-CCD (BI) | HEO | 0.078 |
| XMM | EPIC | MOS-CCD (FI) | HEO | 0.026 |
| SRG | eRosita* | CCD (BI)** | L2 | **0.07** |
| *ATHENA* | X-IFU | Microcalorimeters | L2 | **0.31** |
| *ATHENA*best | X-IFU* | Microcalorimeters | L2 | **0.05** |
| IXO | WFI* | Silicon drift | L2 | **0.01** |

**Notes.** Data in bold are simulated. (*) Detector surrounded by an electron-shielding material. (**) Value obtained without pattern analysis.
**References.** Lotti et al. (2012); Perinati et al. (2011); Lumb et al. (2002); Hauf et al. (2009); Kuntz et al. (2008).

was set to 5%. From the definition of dead time DT $= 1 - e^{-tAF}$, we can see that a DT of 5% is induced by a particle flux above

$$F = -\frac{\ln(0.95)}{tA},$$

$$t_r = Nt_{etf} + t_{th}\log\left(\frac{E_{dep}}{E_{max}}\right),$$

where $t_r$ is the time needed to return to the working point, $A \sim 6.25 \times 10^{-4}$ cm$^2$ is the pixel area, $t_{etf}$ is the electrothermal feedback time constant, $t_{th}$ is the thermal time constant, $E_{dep}$ is the energy deposited and $E_{max}$ is the maximum energy that the detector can absorb without saturating. For the main detector we assumed the high-resolution events case $N = 40$ (Barret et al. 2013) and established $t_r \sim 13$ ms. This value was found assuming that every incident particle deposits $\sim 40$ keV in the detector, bringing it into its saturation regime, from which the return to the working point takes longer, and thus constitutes a very conservative estimate. This way we obtain a flux value on the detector of $F_{int} \sim 6000$ cts cm$^{-2}$ s$^{-1}$ that is generated by an external flux of $F_{ext} \sim 10^4$ cts cm$^{-2}$ s$^{-1}$.

According to the Solar Radiation Environment Monitor (SREM) onboard *Planck*, in the period 2009−2013 this threshold was never reached, so we can conclude that the main detector performances will not be hindered by solar energetic events. This result was to be expected because of the small surface of the pixels and their fast response time.

This result obtained for the main detector is not valid for the anticoincidence system, which exploits larger and thicker absorbers. Repeating the same analysis for the ACD, we assumed $t_r \sim 2$ ms and $A \sim 0.8$ cm$^2$, obtaining a lower threshold value, $F_{ACD} \sim 47$ cts cm$^{-2}$ s$^{-1}$, following the same procedure, translates into a shut-down time of 0.32%. In Fig. 3 we show the shut-down time percentage as function of the L2 proton flux.

In conclusion, we can state that the performances of X-IFU will not be affected by solar particle events.

### 2.2. Considerations on the comparison with other instruments

After calculating the internal background for a microcalorimeter array in different setups/orbits, we compared this with the background levels measured or predicted for different instruments (i.e., CCD cameras or silicon drift detectors). Unfortunately, there are no experimental data regarding the background of X-ray microcalorimeters in space, except from the small dataset taken by *Suzaku* XRS before the cooling system failure. The Planck team reported that the particle background at L2 induced a 15% data loss on the bolometers of the High Frequency Instrument (HFI; Catalano et al. 2014). Although the differences between the instruments do not allow a direct comparison of the two backgrounds, indeed the HFI measurements confirm the absolute relevance of the problem. The comparison with other instruments operating in the soft X-ray band is somewhat difficult, because of the different conditions in which the detectors are placed, the different materials that surround them, and different detector geometries (pixel sizes, thicknesses); but from the background levels reported in Table 1 some general statements can be made.

The background is generated mainly by primary protons crossing the spacecraft and by secondary electrons generated in the structures surrounding the detector. The first component is efficiently discriminated in all the cases considered: in thick detectors like silicon-drift and CCD cameras (generally tens of μm thick) MIPs release energies above the upper threshold of the detector; in thin detectors like microcalorimeters, anticoincidence detectors can be used to discriminate this component, even though the mean energy loss is inside the detector energy band. Furthermore, MIPs often turn on more than one pixel, and can be discriminated using pattern-recognition algorithms.

Generally, secondary electrons are more difficult to discriminate: their spectrum decreases with energy as a power-law and thus the bulk of them is produced at low energies (Geant4 collaboration 2003), and they deposit all their energy in a few μm of silicon. Their role can be influenced by several factors:

– In front-illuminated CCDs (FI) there is a non-sensitive layer (generally Si fractions of μm thick on the rear surface, and the electrodes on the front) that absorbs low-energy secondary electrons. Thus FI cameras exhibit a background rate approximately three times lower than back-illuminated (BI) devices in every considered orbit. Furthermore, the non-sensitive layers absorb some energy of the higher energy electrons, and the result is a shift of the impacting electron spectrum to lower energies. Because of the decreasing shape of the spectrum of the emitted electrons, this further reduces the total background in front-illuminated devices.
- The different orbits provide different environments, and fluxes of secondary electrons. The flux of secondary electrons scales approximately with the flux of the primary protons that creates them, and thus it increases from LEO to HEO to L2. Detectors that do not possess a dead layer (BI CCDs, microcalorimeters, silicon drift) experience the same impacting background level when placed in the same environment.
- The actual background strongly depends on the materials surrounding the detectors and on the efficiency in discriminating non-X-ray events (that in turn depends on the detector geometry and processing software). Table 1 shows that detectors surrounded by electron-shielding materials experience lower background levels.
- Smaller pixel sizes allow higher rejection efficiencies for secondary electrons, since it is more likely that different electrons impact different pixels and that the event will be vetoed by pattern-recognition algorithms (Takayasu et al. 2008).

Given all these factors that influence the detector background, it is hard to compare the background levels measured or simulated in the different detectors. The only data we have from detectors of the same kind, placed in the same orbit and surrounded by similar materials, are BI CCDs and microcalorimeter arrays in LEO orbit, which exhibit the same background level.

We would like to pose particular emphasis on the background predicted for the WFI, since it is the other focal plane instrument onboard *ATHENA*. The background level predicted for X-IFU is a factor of ~5 above the corresponding one reported for the WFI (Hauf et al. 2012), even though the latter is not provided with an anticoincidence system. To understand this difference it is necessary to analyze how the two instruments respond to the different background components described above.

The primary protons are efficiently discriminated in both detectors: in X-IFU by the anticoincidence system with an efficiency of >99%, while in the WFI (~450 $\mu$m thick) MIPs release energies above the upper threshold of the detector or are vetoed by pattern-recognition algorithms (the efficiency of WFI pattern analysis is higher than 99%). The difference then must be in the secondary electrons.

The two detectors are placed in the same environment, so we should expect the same electron flux for X-IFU and WFI, but:

- For the WFI a 2.3 protons cm$^{-2}$ s$^{-1}$ flux has been assumed, which is roughly half the flux reported by CREME96 for the L2 orbit for the solar minimum (Marcus et al. 2012; Tylka et al. 1997; Weller et al. 2010), which is the level we assumed in our work as a conservative value representative of the worst case. In addition, the WFI is better protected than X-IFU by a graded-Z shielding.
- In X-IFU there is a Kapton layer inside the inner cryostat surface that blocks low-energy electrons, while in the WFI the graded Z shielding carries out the same function. The electron spectral shapes are flat for both detectors, pointing out that something is absorbing the low-energy end of the electron spectrum. Moreover, the WFI can also exploit an inactive Al layer on the detector surface (~70 nm thick, blocking electrons below ~1 keV) that further enhances this effect.
- The WFI pixel size is 130 × 130 $\mu$m$^2$, more than five times smaller than X-IFU, and this enhances the rejection efficiency obtained with pattern-recognition algorithms.

While the lower proton flux can account for a factor of ~2 in the background difference, the other factors account for the remaining difference. Their relative impact is under investigation.

## 3. Diffuse background

The soft X-ray background (SXRB) is a diffuse X-ray emission observed in every direction in the 0.1−10 keV energy band. It contains several components: above ~1 keV an absorbed power-law, $\Gamma = 1.41 \pm 0.06$ (Moretti et al. 2009), of extragalactic origin dominates the spectrum, produced by the unresolved emission of several AGNs. As the observing time spent on the same field increases, we can expect this diffuse emission to be resolved into the single point sources of which it consists, and thus a decrease of this contribution.

At energies below ~1 keV line emission from hot diffuse gas becomes important: a first important contribution is given by ~$2 \times 10^6$ K gas in the Galactic halo and absorbed by the interstellar medium. The emission from the halo is thought to be uniform and its flux dependent on the column density of hydrogen along the line of sight. The observations show that there is another component whose intensity does not anticorrelate with $n_H$ and which has to originate nearer to us. To explain this foreground contribution a second component of the SXRB has been proposed, generated by a ~100 pc radius cavity around the solar system filled by hot rarefied interstellar gas with $T \sim 1 \times 10^6$ K, a Local Bubble (LB; Bowyer et al. 1968; Sanders et al. 1877).

Recent results (Koutroumpa et al. 2007) entirely explain the not anticorrelated component in the 3/4 keV band with the solar wind charge exchange (SWCX) mechanism. Charge exchange occurs between heavy ions inside the solar wind (mostly oxygen) and neutral atoms inside the solar system. Oxygen "steals" an electron from H or He, and relaxes from its excited state, emitting a line of the OVII triplet or the OVIII line. The oxygen line emission accounts for ~50−80% of the SWCX emission in the 3/4 keV band (Mc Cammon et al. 2002; Henley & Sheaton 2012). The neutrals can be H and He atoms drifting through the solar system from the point [$l \sim 252°, b \sim 9°$] because of its motion through the Galaxy at ~25 km s$^{-1}$ (heliospheric emission), or atoms from the Earth atmosphere, inside the magnetopause (geocoronal emission). Geocoronal emission can exhibit fast variation, but this is not a problem for X-ray satellites that orbit far from Earth.

### 3.1. Variability

The intensity of all these components can vary with time because of the solar cycle and the positioning of Earth with respect to the He cone (SWCX), and due the direction of the line of sight (SWCX, Galactic halo). Emission from the Galactic halo is generally thought to be uniform in galactic latitude, with emission measure depending on the $n_H$ along the line of sight, although different models with emission measure dependent on the galactic latitude have been proposed and have not been ruled out (Henley & Sheaton 2012).

Heliospheric solar wind charge exchange line intensities directly depend on the neutral H and He concentration and on the metals present in the solar wind, which in turn depend on the solar cycle, ecliptic latitude, and on the Earth location along the orbit (Wargelin et al. 2009), and can exhibit sudden increments due to solar wind variations in intensity and composition and due to the passage of coronal mass ejections (CME).

There is also a seasonal variation in SWCX intensity. The reason for this is that during early June the Earth is upstream of the Sun with respect to the flow of neutrals, while it is placed downstream during early December. In the downwind direction the SWCX is weak since the emission originates from several AU away. In addition, one must take into account that H and He behave differently when entering the heliosphere, because of different sensitivity to radiation pressure. The neutral hydrogen distribution is dictated by the effects of gravity, radiation pressure, UV ionizations, and charge exchange with heavy ions. Helium behaves differently: since radiation pressure is negligible, it only experiences the effects of gravity and forms a focused cone downwind of the Sun. Observations from inside the He cone, or looking through the He cone then experience higher levels of SWCX (Koutroumpa et al. 2009).

With the energy resolution typical of microcalorimeters like X-IFU it is possible to separate the diffuse background into its single components; but with the present instrumentation a general behavior can be derived from the observations:

– Variations with time: the contribution to the low-energy SXRB from the Galactic halo and LB is thought to be quite stable, and any variation of the flux in this band is attributable to the SWCX. Thus multiple observations of the same field at different times should be able to place constraints on the SWCX models. The present models predict that during solar minima a higher SWCX between $\pm 20°$ from the ecliptic is expected, because of the higher concentration of heavy ions. During solar maxima, a higher flux is instead expected from high ecliptic latitudes (Koutroumpa et al. 2007). Observations, show a systematic decrease of the flux from solar maximum to solar minimum however, and this contradicts the expectations at low ecliptic latitudes. A possible explanation is that during solar maxima observations are more likely to be contaminated by CMEs. Variations with time are usually only a fraction of the unenhanced flux, but several cases are recorded of more than an order of magnitude increase in $\lesssim 1$ keV emission (Henley & Sheaton 2012).
– Variations with Galactic longitude: variations of SXRB with longitude and latitude provide information on the Galactic halo component. *XMM-Newton* observations showed that (aside from peculiar emission features and low latitudes that reflect changes in the disk) the emission is higher toward the Galactic center ($l \lesssim 270°$ or $l \gtrsim 90°$) for any latitude. This may indicate that the supernovae rate increases toward the Galactic center (favoring a Galactic fountains origin of the halo), or the presence of accreted gas centered on the Galactic center, or the presence of unresolved discrete sources like low-mass X-ray binaries.
– Variations with Galactic latitude: Henley & Sheaton (2012) reported a comprehensive analysis of the variation of OVII and OVIII line with latitude for $|b| \gtrsim 30°$. Their study reported no global behavior for the halo component with latitude: some regions exhibit a positive correlation between intensity and latitude, some regions show an anticorrelation, and other regions show no significant correlation. Furthermore, OVII and OVIII intensities behave differently in the same zones. For $|b| \lesssim 30°$ there is a flux increment of up to a factor of 6 with respect to high Galactic latitudes, and along the Galactic plane a deep minimum, the "absorption trough" is found in the direction of the Galactic center (Snowden et al. 1997, see Fig. 12a).

**Table 2.** Parameters for the diffuse background model: $apec + wabs * (apec + powerlaw)$.

| Component | Parameter | Unit | Value |
|---|---|---|---|
| apec | $kT$ | keV | 0.099 |
| apec | abundance | | 1 |
| apec | redshift | | 0.0 |
| apec | norm | | $1.7 \times 10^{-6}$ |
| wabs | $n_H$ | $10^{22}$ | 0.018 |
| apec | $kT$ | keV | 0.225 |
| apec | abundance | | 1 |
| apec | redshift | | 0.0 |
| apec | norm | | $7.3 \times 10^{-7}$ |
| power-law | photon index | | 1.52 |
| power-law | Norm[a] | | $1.0 \times 10^{-6}$ |
| power-law | Norm[b] | | $2.0 \times 10^{-7}$ |

**Notes.** [a] Point sources. [b] Extended sources.

Variations with position can be as high as 55% (Henley & Sheaton 2012; Henley et al. 2013) outside the Galactic plane. In conclusion, the background in the $\lesssim 1$ keV band is very dependent on the observation conditions, and improvements of the emission models from Galactic halo, SWCX, and Local Bubble are needed to constrain this component. Furthermore, efforts to constrain the SWCX contribution to the SXRB can be made with on/off observations of dark clouds and multiple observations of the same field in different epochs. Only with an energy resolution typical of microcalorimeters like X-IFU will it be possible to separate the single components of SXRB. The SWCX spectra differ from thermal spectra, with unique signatures such as He-like forbidden and intercombination lines (Wargelin et al. 2009). Identifying these lines is essential for using them as a diagnostic tool for the charge exchange component of the SXRB.

### 3.2. Diffuse background model

To model the SXRB we used the parameters reported in Mc Cammon et al. (2002), extracted from a 1 sr region centered on $l = 90°$, $b = +60$. This sky area avoids emission features, such as the Scorpion-Centaurus superbubble, and should be representative of typical high Galactic latitude pointings. It is modeled inside *XSPEC* as the sum of a thermal emission *apec* model (emission spectrum from collisionally ionized diffuse gas that includes line emissions from several elements) from the halo, and a *powerlaw* model (a simple photon power-law that dominates at higher energies) to represent the high-energy unresolved AGN component, multiplied by a *wabs* model for the photoelectric absorption at low energies, with the addition of an unabsorbed thermal *apec* component from the Local Bubble and SWCX. The model parameters are reported in Table 2 (normalizations refer to 1 arcmin$^2$) and assume different normalizations of the extragalactic component: while in the observation of point sources the unresolved emission must be completely taken into account, in the observation of extended sources AGNs can be mostly resolved. We conservatively assumed that in the observation of diffuse sources the diffuse power-law component can be resolved to up to 80%.

Since the model accounts for geocoronal emission from SWCX (not present in L2) we expect it to be conservative. Moreover, for deep surveys what we are doing is to resolve the background into point sources, and assuming the CXB power-law component to be completely unresolved is a very conservative assumption since the probability to have a source with fluxes similar to the background in the extraction spot is low.

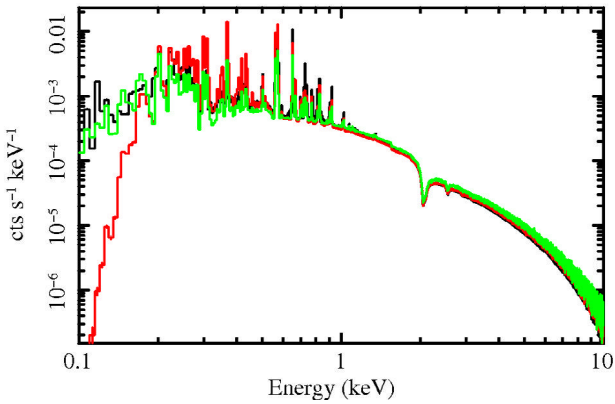

**Fig. 4.** Fluxes produced by the models of McCammon and Lumb for point sources (extraction radius 5″) on X-IFU. In black we plot the McCammon model, in red the Lumb model, and in green our modified version of the Lumb model. Differences in the line intensities above 0.2 keV are due to the choice of different models (*apec* and *mekal* can give different intensities), and to different pointing directions, solar cycle phase, and Earth positioning, as explained in Sect. 3.1.

The predicted fluxes are consistent with the model independently derived by Lumb et al. (2012) in the X-IFU energy range. Our model diverges from that of Lumb at energies below 0.2 keV, because in our model the unabsorbed *apec* component, corresponding to the Local Bubble and SWCX, dominates in this energy range. In the model of Lumb *phabs* $\times$ (*power* + *mekal* + *mekal*) the *mekal* component that accounts for the local emission ($kT \sim 0.1$ keV) is absorbed as the halo component ($n_H = 2 \times 10^{20}$ atoms cm$^{-2}$). If we use a modified version *mekal* + *phabs* $\times$ (*power* + *mekal*) of the Lumb model to fit the data produced by the McCammon model in the 0.2−10 keV energy range, the two agree quite well also below 0.2 keV, as shown in Fig. 4, when one of the *mekal* components is unabsorbed.

## 4. Effects of the background on the observations

Both internal and diffuse components were estimated for the X-IFU instrument in the *ATHENA* configuration and are plotted in Fig. 5 for extended and point sources. It is clear from the figure that the particle background has a relevant impact only for energies above 2 keV (3 keV) for extended (point) sources. Since it is not possible to reduce the diffuse component, obtaining a particle background whose value is lower (hopefully negligible) than the diffuse one is a highly desirable result.

In the following we simulate the observation of two representative cases for the *ATHENA* X-IFU where the low background is particularly crucial, that is, faint and/or diffuse sources. In addition, we quantify the improvement of science performance for these cases. Since our reduced background setup has been adopted as the reference value for *ATHENA*, a full description of the scientific performances and simulations can be found in Nandra et al. (2013). Observations with the nominal background and without the anticoincidence detector have been reported in Lotti (2013). After calculating the total background level it is straightforward to derive the minimum count rates and fluxes obtainable with X-IFU in the different energy bands.

**Table 3.** Limit fluxes for *ATHENA* X-IFU for $t = 100$ ks.

| energy range [keV] | Extended sources (1 arcmin$^2$) | | |
|---|---|---|---|
| | $B_{tot}$ [cts s$^{-1}$ keV$^{-1}$] | $R_{min}$ [cts s$^{-1}$] | $F_{min}$ [erg cm$^{-2}$ s$^{-1}$] |
| 0.2–10 | $4.3 \times 10^{-3}$ | $3.2 \times 10^{-3}$ | $8.2 \times 10^{-16}$ |
| 0.5–2 | $1.2 \times 10^{-2}$ | $2.1 \times 10^{-3}$ | $2.6 \times 10^{-16}$ |
| 2–10 | $8.4 \times 10^{-4}$ | $1.3 \times 10^{-3}$ | $1.8 \times 10^{-15}$ |
| Point sources ($R_{extraction} = 5$ arcsec) | | | |
| 0.2–10 | $1.4 \times 10^{-4}$ | $5.9 \times 10^{-4}$ | $1.5 \times 10^{-16}$ |
| 0.5–2 | $4.7 \times 10^{-4}$ | $4.2 \times 10^{-4}$ | $5.1 \times 10^{-17}$ |
| 2–10 | $2.6 \times 10^{-5}$ | $2.3 \times 10^{-4}$ | $3.2 \times 10^{-16}$ |

In Table 3 and in Fig. 6 we report the values for extended (1 arcmin$^2$) and point (extraction radius 5 arcsec) sources for 100 ks exposure time, assuming the *ATHENA* effective area, a power-law source spectrum ($n_H = 10^{20}$ cm$^{-2}$, $\Gamma = 1.7$), and a $5\sigma$ significance.

### 4.1. Point sources

As a representative case for a point source we have taken the case of a high-redshift Compton-thick AGN. There is strong evidence of an increasing fraction of obscured AGNs at faint fluxes, even though the very steep increase from high to low fluxes is poorly reproduced by the models, suggesting a poor sampling of the Compton-thick AGN population even in the deep surveys with Chandra and XMM. The population synthesis models predict an upturn in the number of Compton-thick AGN at fluxes of $\sim 10^{-15}$ erg cm$^{-2}$ s$^{-1}$ [2−10 keV] (Georgantopoulous et al. 2012), so it is vital to investigate the Compton-thick AGN population around and below such fluxes.

To demonstrate the detection capabilities of X-IFU and the role of background in the observation of point sources we simulated the observation of CXOCDFSJ033229.8-275106 (XID202 in the numbering scheme of Giacconi et al. 2002), a distant, obscured AGN as obtained with the ultra-deep ($\sim$3Ms) *XMM-Newton* survey in the *Chandra* deep field South (Comastri et al. 2011). We scaled the XID202 flux by a factor of 6, to make it fainter and closer to the limit flux for X-IFU, and so probe the *ATHENA* capabilities of characterizing Compton-thick AGNs where the current surveys mostly lack sensitivity. The scaled AGN (hereafter $XID202_{1/6}$) has a flux of $6.38 \times 10^{-16}$ erg cm$^{-2}$ s$^{-1}$ in the $0.5−10$ keV energy band and a luminosity $L \sim 10^{44}$ erg s$^{-1}$.

For the AGN spectrum we used a *zpowerlw* (a redshifted power-law) and a *zgauss* (a redshifted Gaussian line profile) models absorbed by a *zwabs* model. The parameters are given in Table 4. In Fig. 7 we plot the spectra obtained with X-IFU with two different observational times together with the background level.

To fit the simulated spectra we use the *FTOOLS* to rebin the data until each bin had at least 30 counts.

The results of the fit are reported in Table 5. The strong iron line is correctly identified at the 6.38 keV energy in the rest frame and the source redshift $z = 3.7$ is recovered with negligible relative errors, and thus is frozen in the fit. The source count rate is $\sim 1.1 \times 10^{-3}$ cts s$^{-1}$, of which $\sim 4.8 \times 10^{-4}$ cts s$^{-1}$ in the iron line. The background count rate is $\sim 9.3 \times 10^{-4}$, and thus the S/N ratio is 42 for the 3 Ms observation and 9 for the 150 ks observation. Sources at this flux level needs $t \sim 42$ ks to be detected by *ATHENA* with confidence level $> 5\sigma$. To detect the line with $n_\sigma$ significance the equation $\frac{S_{\text{line}}}{\sqrt{S_{\text{line}}+S_{\text{cont}}+B}} \sqrt{t} > n_\sigma$ must be satisfied, where $S_{\text{line}}$, $S_{\text{cont}}$ and $B$ are the source, continuum and background count rates, respectively. The AGN iron $K_\alpha$ line produces $S_{\text{line}} = 4.8 \times 10^{-4}$ cts s$^{-1}$, and since $S_{\text{line}} \ll S_{\text{cont}} \ll B$ for $n_\sigma = 5$ it is easy to find the condition $t > \sim 50$ ks.

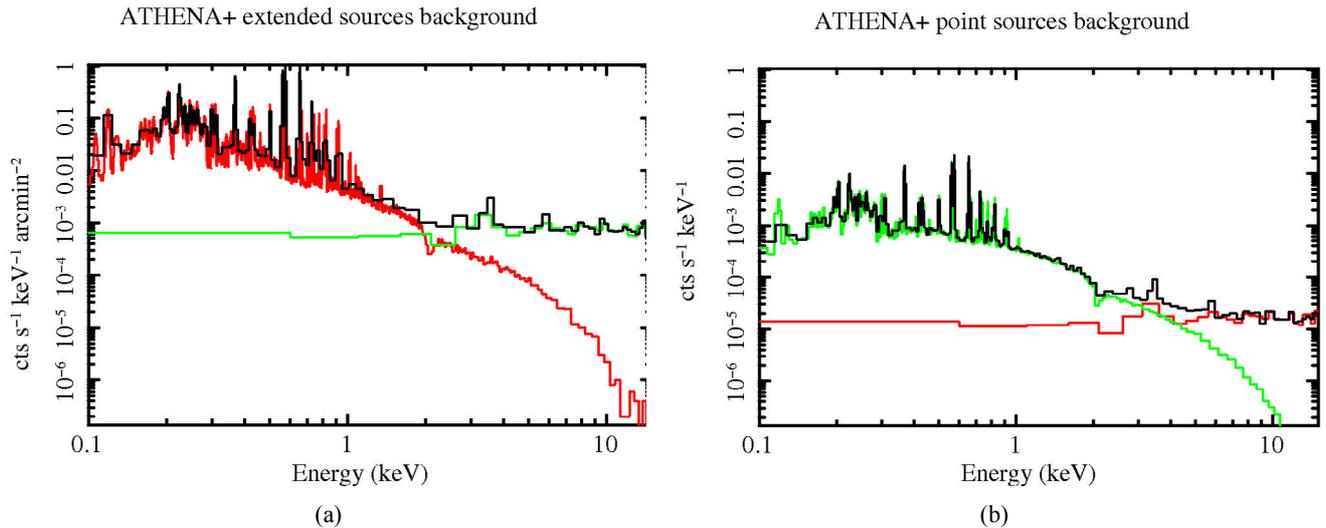

**Fig. 5.** Background components for extended **a)** and point **b)** sources: the green line is the internal particle background calculated with GEANT4, the red line is the sum of the unresolved CXB component and of the Galactic foreground. The total background is the black line. The background files for point sources assume an extraction radius of 5 arcsec.

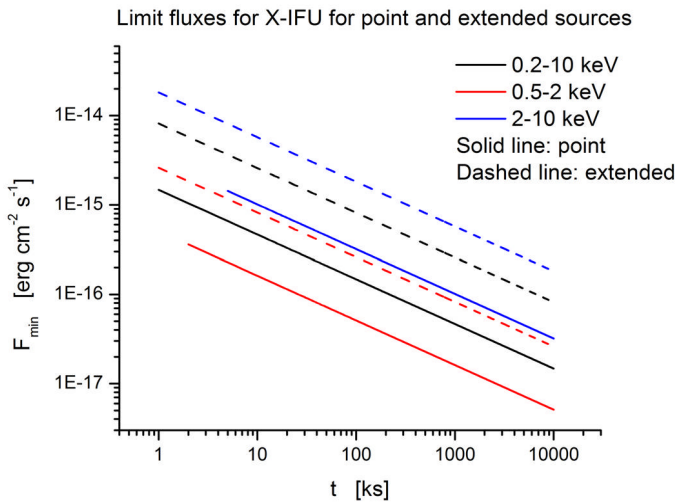

**Fig. 6.** Limit fluxes for *ATHENA* X-IFU as a function of time for point (solid lines) and extended (dashed lines) sources in different energy ranges.

**Table 4.** XID2021,6 model parameters: *zwabs* ∗ (*zpowerlw* + *zgauss*). Flux$_{[0.5-10\text{ keV}]} = 6.4 \times 10^{-16}$ erg cm$^{-2}$ s$^{-1}$.

| Component | Parameter | Unit | Value |
|---|---|---|---|
| zwabs | $n_H$ | $10^{22}$ | 88 |
| zwabs | redshift | | 3.7 |
| zpowerlw | photon index | keV | 1.82 |
| zpowerlw | redshift | | 3.7 |
| zpowerlw | normalization | | $3.17 \times 10^{-6}$ |
| zgauss | line E | keV | 6.38 |
| zgauss | σ | keV | 0 |
| zgauss | redshift | | 3.7 |
| zgauss | normalization | | $6.67 \times 10^{-7}$ |

**Table 5.** Relative errors on hydrogen column densities, power-law index, iron $K_\alpha$-line intensities, and S/N.

| | XID2021,6 | | | | | |
|---|---|---|---|---|---|---|
| $t$ | $\sigma(n_H)/n_H$ | | $\sigma(\Gamma)/\Gamma$ | | $\sigma(I)/I$ | S/N |
| 3 Ms | 4.1% | | 3.3% | | 6.3% | 42 |
| 150 ks | 21% | | 17% | | 32% | 9 |

– *Expected detections*

To estimate the number of the sources that will be detected and characterized by *ATHENA* we extrapolated the luminosity-dependent density evolution (LDDE) model described in La France et al. (2005) out to redshift 7. This way we expect to find 4570 AGN per square degree between $z = 0$ and $z = 7$ above the detection threshold of X-IFU for 100 ks, and if we correct for the Compton-thick fraction reported by Brightman et al. (2012) we obtain 1560 CT ($N_H > 10^{24}$ cm$^2$) AGN/deg$^2$ detectable by X-IFU. Given the 25 arcmin$^2$ FoV of *ATHENA* this translates into ~32 AGN for each 100 ks observation (~11 CT) between $z = 0$ and $z = 7$. If we restrict our investigation to high-redshifts (i.e., $z > 2$) we expect to find ~4.2 AGN/FoV, of which ~1.8 will be CT AGNs.

Assuming that at least 50% of the mission lifetime will be spent in high Galactic latitude surveys with $t > 100$ ks, with a 0.8 observation efficiency (i.e., 12.6 Ms/yr effective observational time), we expect to serendipitously find and characterize ~12 000 new AGNs (~4100 CT) in a three-years mission lifetime between $z = 0$ and $z = 7$, of which ~1600 will be found at $z > 2$ (~660 CT).

### 4.2. Diffuse sources

Extended sources experience higher background levels, because the detector area exposed to them is larger, and thus the lowest detectable flux is higher than that of the point sources, as can be seen in Table 3. Below we simulate the observation of a galaxy cluster, MACSJ1621.3+3810, an extended source at $z = 0.456$ with a brightness of $8.61 \times 10^{-16}$ erg cm$^{-2}$ s$^{-1}$ arcmin$^{-2}$ in the 0.5−2 keV energy band. We used an extraction area

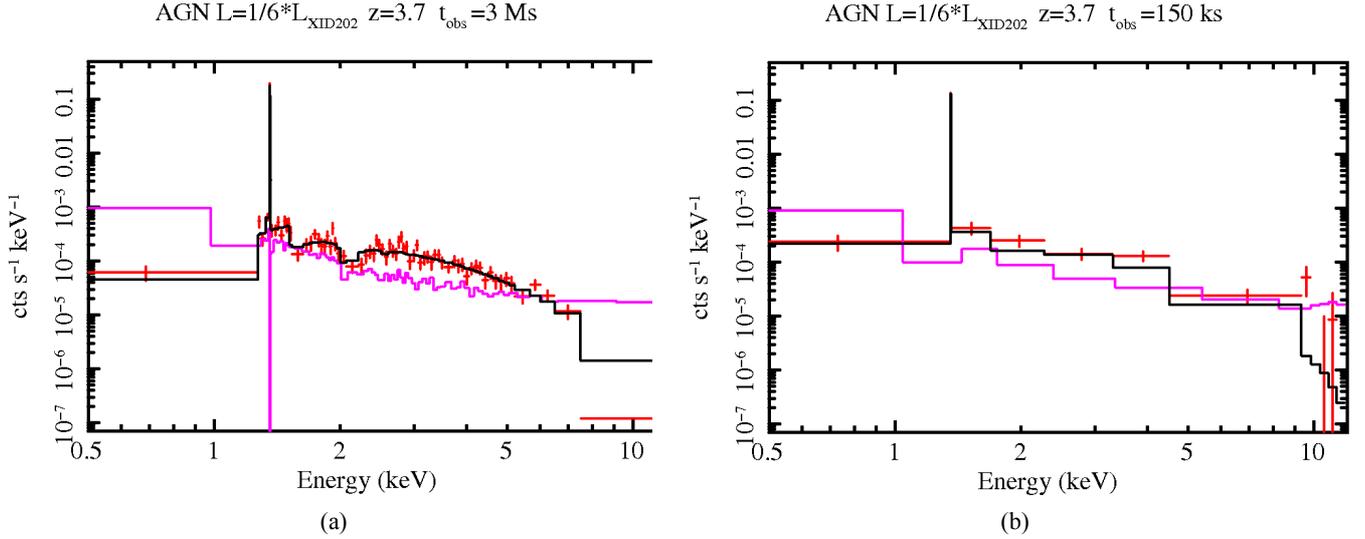

**Fig. 7.** Spectra of XID2021_6 for a 3 Ms **a)** and 150 ks **b)** observations with *ATHENA*. The black line depicts the model, the red crosses the data, and the cyan line delineates the background level. The counts are binned to obtain at least 5$\sigma$ significance.

**Table 6.** MACSJ1621.3 model parameters.

| Component | Parameter | Unit | Value |
|---|---|---|---|
| wabs | $n_H$ | $10^{22}$ | 0.011 |
| apec | $kT$ | keV | 6.62 |
| apec | abundance |  | 0.3 |
| apec | redshift |  | 0.0465 |
| apec | normalization |  | $2.77 \times 10^{-5}$ |

**Notes.** $A_{extr} = 10$ arcmin$^2$, $S^{0.5-2keV}_{MACSJ1621.3} = 8.61 \times 10^{-16}$ erg cm$^{-2}$ s$^{-1}$ arcmin$^{-2}$.

of 10 arcmin$^2$. In modeling the cluster parameters we assumed that the background spectrum is known and that the only parameters to be varied are the cluster parameters.

– *MACSJ1621.3+3810*

To simulate the observation of the MACSJ1621.3+3810 cluster we followed Ettori et al. (2011), and modeled the spectrum using the *apec* model to simulate the emission of the intra-cluster medium, and a *wabs* model for the Galactic H absorption. The cluster parameters at $R_{200}$ are reported in Table 6.

We simulated a 50 ks and a 500 ks observation with *ATHENA*, taking into account the fluctuations of the background, and then subtracted our background model from that file. The results are shown in Fig. 8a. We grouped the data to have at least 30 counts in each bin and performed the spectral fitting to obtain the cluster metallicity, redshift, temperature, and elemental abundances. The reduced background level allowed the iron $K_\alpha$ line to emerge from the background and be identified, while with the previous "nominal" value this was not possible (Lotti et al. 2013).

The procedure used to analyze the data is the following:

– We fixed the Galactic absorption column $n_H$ and the redshift to the input values used to simulate the observation (Table 6).
– We fit the data with an absorbed *apec* model in the 0.2 − 10 keV energy band to obtain the temperature, global metallicity and normalization.

**Table 7.** Relative errors for the 2 observations of MACSJ1621.3 described in the text.

| | | Relative error | |
|---|---|---|---|
| Parameter | | 50 ks | 500 ks |
| $\sigma_T/T$ | $T$ | 16 (21)% | 4.8 (7.6)% |
| $\sigma_\rho/\rho$ | | 2.6 (3.3)% | 0.7 (0.9)% |
| $\sigma_Z/Z$ | $Z$ | 40 (53)% | 12 (16)% |
| $\sigma_{n_O}/n_O$ | $n_O$ |  | 24 (28)% |
| $\sigma_{n_{Ne}}/n_{Ne}$ | $n_{Ne}$ |  | 55 (55)% |
| $\sigma_{n_{Si}}/n_{Si}$ | $n_{Si}$ | 75 (–)% | 24 (25)% |
| $\sigma_{n_S}/n_S$ | $n_S$ |  | 35 (90)% |
| $\sigma_{n_{Ar}}/n_{Ar}$ | $n_{Ar}$ |  | 76 (–)% |
| $\sigma_{n_{Ca}}/n_{Ca}$ | $n_{Ca}$ |  | 81 (82) |
| $\sigma_{n_{Fe}}/n_{Fe}$ | $n_{Fe}$ | 35 (46)% | 11 (12)% |
| $\chi^2_{red}$ | | 1.01 | 0.97 |
| d.o.f. | | 629 | 3503 |

**Notes.** Values in parenthesis refer to the errors obtained with the nominal background, while error values $\geq 100$% are left blank. The last two lines report the reduced $\chi^2$ and the degrees of freedom of the *apec* fit.

– We also fixed the $T$ value found and fit the data again with a *vapec* model in the same energy range to obtain the single elemental abundances.

In the *vapec* model the abundances of the elements that were not tightly constrained by the fit (i.e., relative error $\geq 100$%) were tied to the iron abundance. This allowed us to recover the temperature, metallicity, and abundances of some target elements (Nandra et al. 2013). The results are reported in Table 7 together with the values obtained with the nominal background for comparison. The fit with the *apec* model reported a reduced $\chi^2 = 1.01$ (629 d.o.f.) for the 50 ks observation and a reduced $\chi^2 = 0.968$ (3503 d.o.f.) for the 500 ks observation.

With these internal background levels it is possible to construct radial entropy profiles ($K = k_B T / n_e^{2/3}$), and of the most abundant elements up to $R_{200}$, showing for the first time how elements are distributed in the entire cluster volume. This will unveil the role of AGN feedback in shaping the metallicity profiles, allowing us to distinguish between different sources of metal enrichment: each source of metals (e.g., SNIa, SNcc, AGB stars) synthesizes heavy elements in different proportions, so their relative role can be assessed by using abundance ratio measurements in different epochs (Ettori et al. 2013) to trace the chemical evolution of metals over cosmic time. Thanks to the ability to perform spatially resolved high-resolution spectroscopy for faint sources even in moderate observational times, it will be possible to study the turbulence and bulk motions in clusters, the physics of cluster accretion in outskirts to constrain models, and to investigate the scale of clumping in outskirts and the clump support mechanism. Improved measurements of cluster masses to $R_{200}$ also allow to understand cluster mass function, which depends on cosmological parameters.

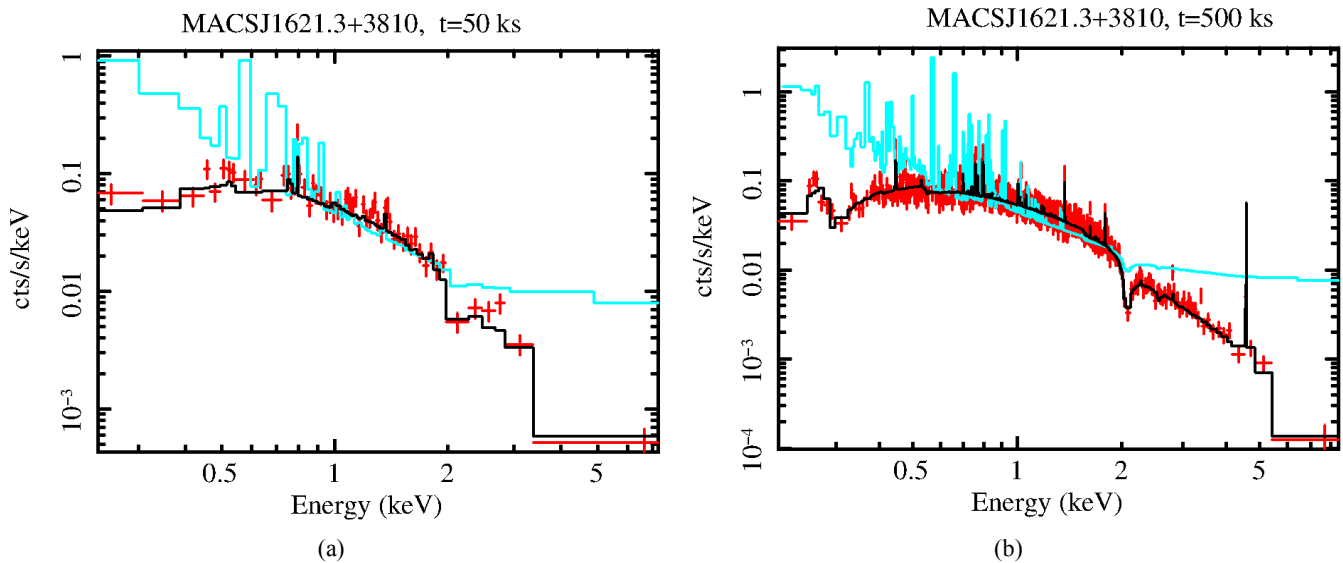

**Fig. 8** Spectra of the 50 ks **a)** and 500 ks **b)** observations of MACSJ1621.3+3810 with ATHENA. The black line depicts the model, the red crosses the data, and the cyan line shows the background level.

## 5. Results and discussion

Thanks to several design solutions, we were able to reduce the particle background level by a factor of 6 with respect to the nominal configuration, and by a factor of 60 with respect to the expected baseline value for L2. The best background level achievable in the L2 orbit resulted similar to the corresponding LEO background, allowing us to exploit the advantages of the L2 orbit without the increased background. This includes stable background levels and thermal environment, avoiding space debris, geomagnetically trapped particles, steep temperature gradients, and occultation by Earth. The internal background level is consistent with a constant value of 0.05 cts cm$^{-2}$ s$^{-1}$ in the 0.2−10 keV energy band. The official background files[3] are the same as adopted in this paper and are based on this constant internal background.

We compared the different internal background levels measured by soft X-ray instruments and found that the internal background is induced mainly by primary protons crossing the spacecraft and by secondary electrons generated in the structures surrounding the detector. The first component is always efficiently discriminated, since in thick detectors MIP release energies above the upper threshold of the detector, while in thin detectors in which the mean energy released falls inside the detector energy band, anticoincidence detectors can be used to discriminate this component. Furthermore, MIPs often turn on more than one pixel and can be discriminated using pattern-recognition algorithms. The secondary electrons are more difficult to discriminate, since they are absorbed in a few $\mu$m of silicon, and their role is influenced by several factors: their flux scales approximately with the flux of the primary protons that creates them and thus increases from LEO to HEO to L2. However, the actual background experienced also strongly depends on the materials surrounding them and on the efficiency in discriminating non-X-ray events, which is dictated by the detector geometry and the processing algorithms, as explained in Sect. 2.2. As a consequence, FI cameras exhibit a background rate approximately three times lower than back-illuminated (BI) devices in every considered orbit, because of their lack of an inactive layer that absorbs secondary electrons, and generally detectors surrounded by electron-shielding materials experience lower background levels.

Given all these factors that influence the detector background, it is hard to compare the measured/simulated background levels in the different detectors. The only data we have from detectors of the same kind, placed in the same orbit and surrounded by similar materials, are BI CCDs and microcalorimeter arrays in LEO orbit, that in fact exhibit the same background level.

We defined a reference model for the diffuse background and assembled all the available informations on its variation with epoch and pointing direction. We found that the diffuse background component strongly depends on the pointing direction and epoch. To constrain this component either an improvement of the current models or dedicated observations with ~eV energy resolution are needed. We also investigated the role of high-energy solar particle events in hindering the detector performances, and concluded that solar flares will not induce appreciable degradation of the X-IFU operational efficiency.

After the foreseen background level was established, it was possible to predict the mission performances in the observation of faint and/or diffuse sources. We analyzed two key cases: the detection of obscured AGNs at high-redshift, and the characterization of diffuse sources such as clusters of galaxies at $R_{200}$.

Active galactic nuclei are thought to be formed at early cosmic times, and efficiently fed at Eddington rates because of the wide availability of dust and gas in the early cosmic environment. If this picture holds, an increase of heavily obscured object is expected with increasing redshift. The current deep X-ray surveys have already sampled the whole unobscured population of Compton-thin objects, being able to correctly reproduce their contribution to the X-ray background (Worsley et al. 2005). However, the properties and evolution of the population of Compton-thick objects are characterized mainly based on how the models reproduce the X-ray background excess at ∼30 keV. There is strong evidence of an increasing fraction of obscured objects at faint fluxes that are poorly reproduced by the models, suggesting an inadequate sampling of the Compton-thick AGN population even in the deep surveys with *Chandra* and XMM. The population synthesis models predict an upturn in the number of Compton-thick AGN at fluxes of $\sim 10^{-15}$ erg cm$^{-2}$ s$^{-1}$ [2−10 keV] (Georgantopoulous et al. 2012). We demonstrated that with this background level the *ATHENA* mission is able to efficiently sample the Compton-thick AGN population below this critical flux value, detecting ∼4100 new obscured AGNs with $F \gtrsim 6.4 \times 10^{-16}$ erg cm$^{-2}$ s$^{-1}$ during the mission lifetime (∼660 at $z \gtrsim 2$), improving by an order of magnitude the S/N ratios for the detection with respect to current facilities such as XMM. Observations with the reduced background value also allow us to improve the times needed for the detection and the characterization of high-redshift and/or faint Compton-thick AGNs by a factor of ∼2 (Lotti et al. 2013).

---
[3] Available from http://www.the-athena-X-ray-observatory.eu/

Clusters of galaxies trace the large-scale distribution of baryonic matter in the Universe. However, the complete history and mechanism of their growth and metal enrichment is still to be fully comprehended. These structures are assumed to be undergoing strong energetic activity around the virial radius as the baryons are accreted into the dark matter potential. Current observations of local systems are limited to relatively bright objects and barely reach $R_{500}$. X-IFU will allow us to characterize cluster of galaxies near $R_{200}$ on timescales of 50 ks (500 ks) with errors ∼40% (∼12%) on metallicity, ∼16% (∼4.8%) on temperature, ∼5.2% (∼1.4%) on the *apec* normalization, which propagates in ∼2.6% (∼0.72%) on the gas density since $\rho \propto \sqrt{norm}$, and several single-element abundances.

The background level obtained enhances the characterization of these extended sources in terms of metallicity, temperature, and single-elemental abundance determination, also allowing us to determine more elemental abundances than with to the nominal design (see also Lotti et al. 2013). Thus spectra obtained with X-IFU will provide gas density and temperature profiles, that turn into entropy and mass profiles, which are vital for localizing where and how accretion takes place. The dependence of relative abundances of different elements with radius also provide insight into the different metal-enrichment mechanisms, their timescales, and how the gas is mixed by gas-dynamical processes, disentangling the contributions of the several mechanisms at work that can transport the metals from the galaxies where these elements were produced into the hot intra-cluster medium. Abundance studies up to the epoch of cluster formation will give unprecedented insight into enrichment process, allowing us to constrain the origin of metals through studying the evolution of relative abundances.

*Acknowledgements.* This work was supported by ASI contract I/035/10/0-10-308. We wish to thank Memmo Federici and Bruno Martino for the development of GFORC (Federici et al. 2011), the computer cluster that allowed the Geant4 simulations in a reasonable time, and Paul Buehler for having guaranteed access to the SREM data and for the help with the analysis. We also wish to thank the referee for his valuable comments which helped to improve the paper.